\documentclass[letterpaper,twocolumn,10pt]{article}
\usepackage{usenix,epsfig,endnotes}
\usepackage{xcolor}
\usepackage{fontawesome}
\usepackage{multirow}
\usepackage{tabularx}
\usepackage{makecell}
\usepackage{enumitem}
\usepackage{listings} 
\usepackage{subcaption}
\usepackage{stfloats} 
\usepackage{pifont}
\usepackage{tikz}
\usepackage[capitalise,nameinlink]{cleveref}

\begin{document}

\date{}

\title{\Large \bf RoCE BALBOA: Service-enhanced Data Center RDMA for SmartNICs}

\author{
{\rm Maximilian Jakob Heer, Benjamin Ramhorst, Yu Zhu, Luhao Liu, Zhiyi Hu, Jonas Dann, Gustavo Alonso}\\
\{maximilian.heer, benjamin.ramhorst, yu.zhu\}@inf.ethz.ch \\ \{luhliu, zhiyihu\}@student.ethz.ch \\\{jonas.dann, alonso\}@inf.ethz.ch\\
ETH Zurich
}

\maketitle

\thispagestyle{empty}

\subsection*{Abstract}
Data-intensive applications in data centers, especially machine learning (ML), have made the network a bottleneck, which in turn has motivated the development of more efficient network protocols and infrastructure. 
For instance, remote direct memory access (RDMA) has become the standard protocol for data transport in the cloud as it minimizes data copies and reduces CPU-utilization via host-bypassing. Similarly, an increasing amount of network functions and infrastructure have moved to accelerators, SmartNICs, and in-network computing to bypass the CPU. 
In this paper we explore the implementation and deployment of RoCE BALBOA, an open-source, RoCE v2-compatible, scalable up to hundred queue-pairs, and 100G-capable RDMA-stack that can be used as the basis for building accelerators and smartNICs. 
RoCE BALBOA is customizable, opening up a design space and offering a degree of adaptability not available in commercial products. We have deployed BALBOA in a cluster using FPGAs and show that it has latency and performance characteristics comparable to commercial NICs. 
We demonstrate its potential by exploring two classes of use cases. One involves enhancements to the protocol for infrastructure purposes (encryption, deep packet inspection using ML). The other showcases the ability to perform line-rate compute offloads with deep pipelines by implementing commercial data preprocessing pipelines for recommender systems that process the data as it arrives from the network before transferring it directly to the GPU. These examples demonstrate how BALBOA enables the exploration and development of SmartNICs and accelerators operating on network data streams.  

\section{Introduction}
The rise of data-intensive applications such as large scale data analytics and ML, has turned the network into a bottleneck \cite{network_bottleneck_training}. In search for higher efficiency, \textit{Remote Direct Memory Access} (RDMA) has emerged as the standard transport protocol in the cloud, accounting for up to 70\% of all network traffic \cite{azure_rdma}. RDMA offloads both network processing and memory access to the NIC and thus combines low latency, high throughput, and low CPU utilization. These properties facilitate, e.g., memory disaggregation - the separation of compute nodes from shared pools of memory with low latency access through RDMA- as a possible design for datacenters \cite{memory_disaggregation_survey, memory_disaggregation_rdma_as_enabler, memory_disaggregation_rdma_as_abstraction}. Being adopted from on-premise HPC systems with vastly different use cases \cite{RDMA_HPC_to_Cloud}, RDMA has nevertheless some known pitfalls in the public cloud: the combination of lacking access control and weak or missing cryptographic support can be exploited as an attack vector \cite{redmark_rdma_weakness, simpson_rdma_pitfalls}.

At the same time, a unique shift in the data center hardware landscape has become noticeable: Due to the end of Moore's law and Dennard scaling, domain-specific architectures (DSA) \cite{golden_age_computer_architecture} in the form of increasingly specialized accelerators such as GPUs, TPUs, DPUs, and FPGAs are on the rise. The devices are playing an important role in addressing the network bottleneck and in moving computation closer to the data through in-network computing in SmartNICs and DPUs:  AWS Nitro \cite{aws_nitro_nic}, Nvidia BlueField \cite{bluefield_dpu}, Meta FBNIC \cite{meta_fbnic}, and the custom Microsoft MANA-NIC \cite{microsoft_azure_mana} are examples of ASIC-based platform. In this context, FPGAs 
have been used to explore the space, build prototypes of bump-in-the-wire-accelerators or SmartNICs, and even been deployed in academic research \cite{fpga_supernic, fpga_smartnic_nn, fpga_smartnic_distributed_ai, fpga_smartnic_kv} and industry \cite{azure_fpga_nic, azure_fpga_cloudscale, microsoft_azure_boost}. 

In this paper, we aim to combine these two trends and provide a better infrastructure for exploring and exploiting RDMA based systems covering both protocol enhancements and function offloads for  data-intensive applications. Therefore, we propose RoCE BALBOA, an open source\footnote{https://github.com/fpgasystems/fpga-network-stack/tree/9eda6ce9a55c0761ee9e66d1eba38ad5c9474aa9}, 100G-capable and fully RoCE-v2 compatible network stack. 
We demonstrate the potential of RoCE BALBOA by deploying a prototype over FPGAs and developing and evaluating its on-stream data processing capabilities with direct access to GPUs, as well as protocol enhancements addressing the limitations of RoCE V2. 


With this paper, we make the following contributions:
\begin{enumerate}[itemsep=2pt, parsep=0pt]
    \item We introduce RoCE BALBOA, an open source, 100G capable, switched network-tested and fully RoCE-v2 compatible RDMA stack for FPGAs that offers direct memory access to server-grade CPUs and GPUs. 
    The stack, deployable on data center FPGAs, offers performance comparable to that of commercial NICs both in throughput and latency while offering flexibility for extensions and modifications. 
    \item We demonstrate how to leverage this extensibility by improving the RoCE v2-protocol with AES-encryption and machine learning-based deep packet inspection on the datapath as examples of line-rate functional enhancements to the protocol.
    \item We illustrate the performance benefits of on-datapath acceleration using SmartNICs with a concrete use case. We deploy a commercial ML-preprocessing pipeline for recommender models \cite{yu_preprocessing_pipelines} on the RDMA stream datapath capable of directly forwarding the results to the GPU without CPU involvement via DMA. 
\end{enumerate}

\section{Background and Motivation}
This section provides a closer analysis of the RDMA protocol and motivates the need for an extensible stack that can be used in SmartNICs for both improving on the protocol and to directly support applications. 

\subsection{RDMA Design Principles and Pitfalls}

The main characteristic of RDMA is offloading network management and memory interaction to the NIC, which interacts directly with the host memory while bypassing the host OS and CPU via DMA over PCIe \cite{roce_performance_evaluation}. Flows between remote nodes in RDMA networks are managed via Queue Pairs (QP) as connection abstraction and used for remote read and write accesses through either one- (\texttt{RDMA WRITE} and \texttt{RDMA READ}) or two-sided (\texttt{RDMA SEND} and \texttt{RDMA RECEIVE}) transactions \cite{infiniband2014roce, infiniband_verb_tutorial}. BALBOA concentrates on one-sided operations as it is most commonly used option  for on-datapath offloads and minimal CPU-involvement as shown in previous works on the management of disaggregated memory \cite{2_1_ford_disaggregated_persistent_memory} and remotely accessed data structures \cite{2_1_disaggregated_tree_indices, 2_1_far_memory_data_structures, 2_1_RDMA_hashing, 2_1_sync_disaggregated_data_structures}. RoCE v2 is the adaptation of RDMA for Ethernet fabric, which BALBOA implements in the Reliable Connection (RC) mode.  
RDMA originated in high performance computing (HPC) environments which have very different characteristics than cloud computing. Thus, while the combination of zero-copy data transfer and host bypassing results in reductions in CPU utilization, lower latency, and higher throughput, it also creates security issues:
\vspace{-6pt}
\begin{itemize}[itemsep=2pt, parsep=0pt]
    \item Bypassing the operating system (OS) implies bypassing all OS-residing security features and access control mechanisms. 
    \item Without direct support for encryption, the RoCE-v2 protocol exposes both headers and payloads to side-channel attacks, and compromises data privacy in the cloud \cite{simpson_rdma_pitfalls}. 
    \item Previous work has outlined the insufficiency of RoCE-specified security mechanisms (rkeys, PSN-checks etc.) and demonstrated how simple it is to hijack and exploit existing RDMA-flows \cite{redmark_rdma_weakness, NVME_RDMA_Weaknesses}. 
\end{itemize}
\vspace{-6pt}
In this paper, we argue that this missing functionality is best added to the protocol in a SmartNIC so as to maintain the advantages of CPU/OS bypass. This proposed direction is backed up by previous work which gives insight on extensions of the protocol for more efficient memory disaggregation \cite{nic_scale_up_mem_disaggregation} or envision such additions for more security \cite{simpson_rdma_pitfalls}. 
It is also important that the RDMA network stack is open so the functionality can be adapted given that many of the components needed are still evolving or even application-specific (e.g., encryption or ML-based packet inspection tools). 

\subsection{The Case for Application Logic Offload}
As the amount of data to be processed grows, the growing gap between computing capability and transport bandwidths becomes more acute, especially in the cloud where storage disaggregation means that all I/O actually involves the network. At the same time, network virtualization is expensive and consumes a lot of CPU cycles \cite{ms_catapult_nic}, which has led to offloading the support for networking away from the CPU into specialized accelerators. In such scenarios, it makes sense to explore the possibility of also offloading application level logic to the NIC \cite{in_network_computing_dumb_idea}, with recent examples ranging from data analytics \cite{HotOS_In_Network_Analytics} to ML preprocessing tasks \cite{survey_in_network_computation}. Existing systems leverage both experimental reconfigurable network-attached accelerators \cite{facl_fpga_nic, honeycomb_fpga_nic} and commodity SmartNICs and DPUs, which increasingly integrate off- and on-datapath processing capabilities. The combination of SmartNIC and GPU, which enables ML-related use cases, such as offloaded object storage clients \cite{ASPLOS_Object_Storage_Client_DPU} or shared memory implementations \cite{gpu_direct_openshmem}, is of particular interest these days. However, as explained in \Cref{subsec:Existing RDMA-solutions}, these GPU-direct solutions are either limited by the comparatively weak off-datapath compute capabilities of commodity NICs and DPUs \cite{Off-Datapath-NIC-Study} or do not offer RDMA-support \cite{zeke_gpu_nic}, making them unsuitable for cloud environments.

In this paper, we demonstrate the advantages of an open and flexible architecture when offloading a ML preprocessing pipeline for recommender systems on the datapath from the RDMA stack to the GPU. This allows data to be moved directly to the memory of the GPU from the network (e.g., from disaggregated storage) while processing it as it passes through the NIC. By doing this, we get the advantages of network offloading and host bypassing while also having the ability to process data at line rate on a platform more suitable to the task than either the GPU or the CPU. The design serves as a blueprint for other SmartNICs and, being open sourced, opens up many possibilities for exploring this rich space.


\section{Related Work}
\label{subsec:Existing RDMA-solutions}

\begin{table*}[hbt!]
\scriptsize
\centering
\caption{Feature overview and comparison of various network solutions, comprising both academic and commercial FPGA stacks, commercial ASIC-based RDMA NICs and the presented BALBOA design as benchmark.}
\renewcommand{\arraystretch}{1.2}
\begin{tabular}{lccccccccc}

 System &
   \begin{tabular}[c]{@{}c@{}}RoCE-v2 \& \\ NIC-compatible\end{tabular} &
   100G-capable &
   Open source &
   \begin{tabular}[c]{@{}c@{}}Works with \\ switches \end{tabular} &
   \begin{tabular}[c]{@{}c@{}}Customizable \\ network stack\end{tabular} &
   {\begin{tabular}[c]{@{}c@{}}DMA to \\ host memory\end{tabular}} &
   {\begin{tabular}[c]{@{}c@{}}DMA to \\ GPU\end{tabular}} &
   \begin{tabular}[c]{@{}c@{}}On-datapath \\ acceleration\end{tabular} &
   \begin{tabular}[c]{@{}c@{}}Protocol \\ enhancements\end{tabular} \\ \hline
 PROP \cite{prop_rdma} &
  \faSquareO &
  \faSquareO &
  \faSquareO &
  \faCheckSquare &
  \faCheckSquare &
  \faCheckSquare &
  \faSquareO &
  \faSquareO &
  \faSquareO \\
 Mansour \cite{mansour_rdma} &
  \faPlusSquareO &
  \faCheckSquare &
  \faSquareO &
  \faSquareO &
  \faCheckSquare &
  \faCheckSquare &
  \faSquareO &
  \faSquareO &
  \faSquareO \\
 StRoM \cite{strom_smart_remote_memory} &
  \faSquareO &
  \faCheckSquare &
  \faCheckSquare &
  \faSquareO &
  \faCheckSquare &
  \faCheckSquare &
  \faSquareO &
  \faCheckSquare &
  \faSquareO \\
 Schelten \cite{schelten_rdma_1, schelten_rdma_2} &
  \faCheckSquare &
  \faCheckSquare &
  \faSquareO &
  \faCheckSquare &
  \faCheckSquare &
  \faSquareO &
  \faSquareO &
  \faCheckSquare &
  \faSquareO \\
 Sun et al. \cite{sun_rdma} &
  \faCheckSquare &
  \faCheckSquare &
  \faSquareO &
  \faSquareO &
  \faCheckSquare &
  \faCheckSquare &
  \faSquareO &
  \faSquareO &
  \faSquareO \\ \hline
 ETRNIC \cite{ETRNIC_RDMA} &
  \faSquareO &
  \faCheckSquare &
  \faSquareO &
  \faSquareO &
  \faSquareO &
  \faSquareO &
  \faSquareO &
  \faPlusSquareO &
  \faSquareO \\
 \begin{tabular}[c]{@{}l@{}}ERNIC \cite{ERNIC_RDMA, RecoNIC_RDMA} \end{tabular}  &
  \faCheckSquare &
  \faCheckSquare &
  \faSquareO &
  \faSquareO &
  \faSquareO &
  \faCheckSquare &
  \faSquareO &
  \faCheckSquare &
  \faSquareO \\ \hline
 \begin{tabular}[c]{@{}l@{}}ConnectX-5 \cite{Mellanox_Connect_X_5_RDMA} \end{tabular} &
  \faCheckSquare &
  \faCheckSquare &
  \faSquareO &
  \faCheckSquare &
  \faSquareO &
  \faCheckSquare &
  \faSquareO &
  \faSquareO &
  \faSquareO \\
 \begin{tabular}[c]{@{}l@{}}BlueField 3 \cite{BlueField_RDMA}\end{tabular} &
  \faCheckSquare &
  \faCheckSquare &
  \faSquareO &
  \faCheckSquare &
  \faSquareO &
  \faCheckSquare &
  \faCheckSquare &
  \faPlusSquareO &
  \faPlusSquareO \\ \hline
  \begin{tabular}[c]{@{}l@{}}RoCE \\ BALBOA\end{tabular} &
  \faCheckSquare &
  \faCheckSquare &
  \faCheckSquare &
  \faCheckSquare &
  \faCheckSquare &
  \faCheckSquare &
  \faCheckSquare &
  \faCheckSquare &
  \faCheckSquare
\end{tabular}
\label{table:comparison_rdma_stacks}

\medskip

\faCheckSquare~- satisfied, \faPlusSquareO~- partially satisfied, \faSquareO~- not satisfied
\end{table*}

RDMA has been extensively used in HPC and, since adoption in data centers, to implement a number of advanced systems such as distributed key value stores \cite{rdma_key_value_storage}, disaggregated memory \cite{rdma_disaggregated_memory_2}, data analytics platforms \cite{rdma_data_analytics}, etc. That work is orthogonal to our efforts and can benefit from the systems we propose. For this reason, in this section, we focus solely on RDMA on SmartNIC implementations and efforts that overlap with the use cases we explore in this paper.

On the commercial side, there are several products supporting RDMA. The Mellanox ConnectX-5 \cite{Mellanox_Connect_X_5_RDMA} is an example of a standard 100G-capable NIC with a full implementation of the network protocol. A SmartNIC version of the system is Nvidia's BlueField 3 \cite{BlueField_RDMA},  combining the network functionality with both compute cores for on- and off-datapath offload as well as hardened cores for encryption and decompression. Such commercial systems are useful, but limiting: The programmable on-datapath offloads are limited by access to cache and memory \cite{Off-Datapath-NIC-Study}; the hardened decompression core fails to meet line rate throughput while lacking support for compression \cite{bluefield_throughput}. Being a closed system, there is no way to either extend the system or address its limitations. This makes it difficult to explore which logic could be offloaded and how to best implement it since the bottlenecks (like memory access, packet transfers to the cores, lack of line rate capabilities) are inherent to the platform and cannot be addressed. 

There are also several implementations of RDMA using FPGA-based platforms, from both academia and industry, and dedicated ASIC-based NICs. We have summarized these systems in 
\Cref{table:comparison_rdma_stacks}, which compares existing solutions across network capabilities, open-source availability, and the support for on-datapath acceleration and communication with GPUs.

One of the earliest approaches to RDMA on FPGAs \cite{prop_rdma} adopts the general principles of RDMA data transport with NIC-offloaded network operations and direct access to host buffers.
However, it neither follows the actual protocol nor achieves 100G line rate. 
Thus, the usefulness in data centers is very limited. 
There are also solutions developed for very concrete use cases (e.g., high-energy physics) \cite{mansour_rdma} that achieve 100G throughput but are not open source, and their intended use case is embedded computing rather than the cloud or datacenters. 
For that purpose, several other solutions have been developed \cite{strom_smart_remote_memory, schelten_rdma_1, schelten_rdma_2}. \cite{strom_smart_remote_memory} is open-source, supports 100G, and has been used extensively for RoCE-enabled FPGA projects \cite{coyote_v1_paper, enzian_paper, tNIC_paper, edm_paper, coyote_v2_paper}, but misses key features for datacenter utilization such as compatibility with commodity NICs and switched networks. 
\cite{schelten_rdma_1, schelten_rdma_2} provide a full RDMA implementation but are not open source and, thus, not extensible. This makes it difficult to include critical missing functionality for realistic datacenter applications such as support for \texttt{RDMA READ} or DMA to the host. Especially the latter restriction limits the applicability of this stack to stand-alone network accelerators and prevents the utilization as a NIC. Both \cite{strom_smart_remote_memory} and \cite{schelten_rdma_1, schelten_rdma_2} explore offloading of application logic to the FPGA as we do here, but are limited due to the missing GPU-DMA-compatibility and combination of service- and function-offloads. 
Some commercial implementations have evolved as the use cases change. The original Xilinx implementation of RDMA \cite{ETRNIC_RDMA} targeted mobile applications; the current version, ERNIC \cite{ERNIC_RDMA}, is more suitable for distributed applications with 100G throughput. In \cite{RecoNIC_RDMA}, it was used to implement on-datapath offloads. However, the stack is oriented towards FPGA-to-FPGA communication and it is closed source, not allowing for direct changes of the packet processing logic. 

The RoCE BALBOA stack aims to provide a more comprehensive coverage of functionality, compatibility with switches, and NICs used in data centers, as well as opening up the possibility of extending the protocol and adding application offloading logic combined with direct access to GPU-memory. 
In doing so, it offers the research community a versatile, data center-compatible and open source platform for deployment and evaluation of network functionality and application offloading.

\section{RoCE BALBOA Architecture}

\begin{figure*}[t]
    \centering
    \includegraphics[width=\textwidth]{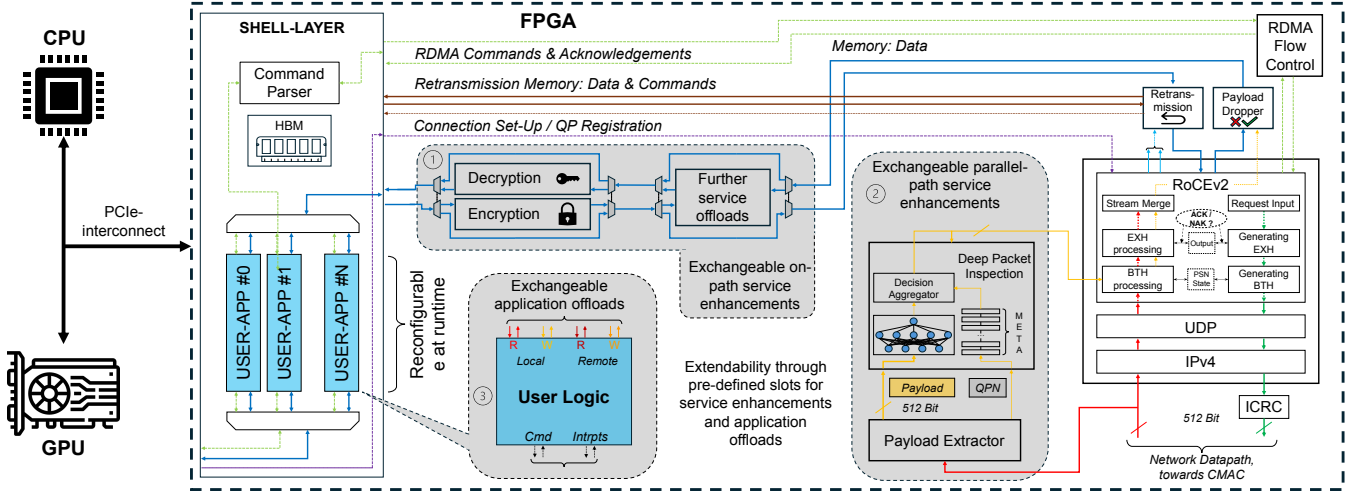}
    \caption{Overview of RoCE BALBOA, consisting of the customizable 100G RoCE packet processing pipelines and stack-adjacent slots for exchangeable on-path (Example: \ding{192} - AES ECB crypto core) and parallel-path service enhancements (Example: \ding{193} - ML-based deep packet inspection). Allows for direct memory access via PCIe to both CPUs and GPUs with data routed through reconfigurable user logic slots \ding{194} in the FPGA shell layer.}
    \label{fig:wide-image}
\end{figure*}

RoCE BALBOA consists of several distinct building blocks for pipelined header processing, checksum calculation, retransmission buffering and flow control (cf. \Cref{fig:wide-image}). All these modules are connected to each other and to the surrounding infrastructure via AXI stream busses for the data and control plane. On the one hand, the combination of 512bit-wide data busses and a system wide clock frequency of 250~MHz allows for an end-to-end throughput of 100G with some over-provisioning for overhead (250~MHz $\times$ 512bit = 128~Gbps). On the other hand, this widely adopted bus standard ensures easy portability of RoCE BALBOA to different FPGA infrastructure designs. While this paper evaluates RoCE BALBOA in the open-source Coyote v2 shell \cite{coyote_v2_paper} with the XDMA \cite{XDMA_IP_Guide} and 100G CMAC \cite{CMAC_IP_Guide} IP cores, a port to the AMD-provided RecoNIC / ONIC-shell \cite{RecoNIC_RDMA} has been realized in the past \cite{maroman_balboa_reconic}.
Furthermore, the utilization of simple AXI streaming interfaces allows the easy integration of further pipelined processing blocks, as demonstrated with the on-datapath services for encryption as well as with the ML-preprocessing pipelines. 

\subsection{Packet Processing Pipeline}
\begin{figure}[t]
    \centering
    \includegraphics[width=0.9\columnwidth]{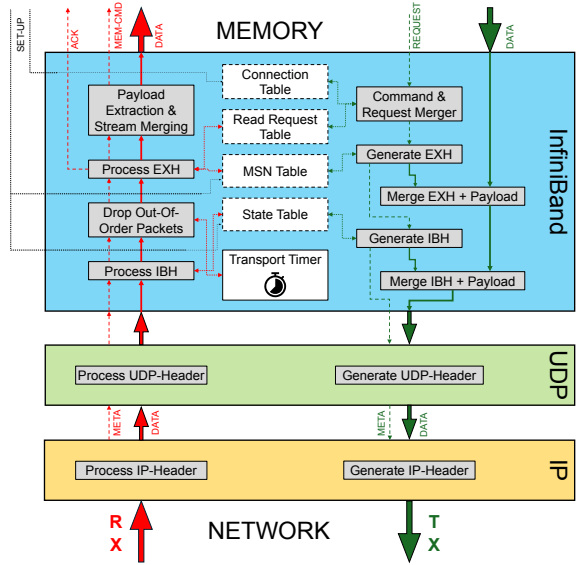}
    \caption{Packet processing pipeline for the sequence of IP, UDP, and InfiniBand headers of RoCE-v2 packets.}
    \label{fig:Packet Processing Pipeline}
\end{figure}

The core of the network stack is the packet processing pipeline implemented in AMD Vitis HLS for high-level configurability (\Cref{fig:Packet Processing Pipeline}). Its structure follows the header stack of RoCE v2 packets, consisting of IP, UDP, and InfiniBand headers and provides a clear separation of RX- and TX-path for incoming and outgoing network packets to allow bidirectional line-rate processing. Both paths share access to state information stored on a per-QP-basis in the multiple tables that are initially configured as part of the QP-setup: The connection table holds high-level information for the remote IP address and UDP port, while the state table keeps track of the packet sequence numbers (PSNs) for a RDMA flow, allowing for checks of PSN duplicates and sequence errors. Finally, the MSN table enables sequence control on a more fine-grained level for multi-message transactions, as it occurs for large buffer transmissions across multiple packets. Per default, these tables support up to 500 QPs, but can be configured for more or less depending on resource restrictions and use case. The transport timer is an additional building block of the processing pipeline and allows for the determination of packet timeouts to trigger retransmissions. 

On the RX-datapath, the respective headers are stripped from the incoming packets, evaluated by finite state machines (FSMs) in the responsible pipeline stage and then converted into meta-commands in parallel control flow busses towards further processing steps. The remaining packet is then realigned before it is handed over in the same direction. 
For the actual RDMA headers, the processing logic is a two-fold process with distinct treatment of the base transport headers (BTH) and the RDMA extended transport headers (RETH): For BTH processing, mainly the RDMA opcode, the QPN, and PSN are extracted from the incoming packet and used for a comparison with the information stored in the state table: By retrieving the existing state of information, the processing logic can determine whether the received PSN equals the expected value and either forward or drop the packet in question based on this comparison. The RETH processing then finally produces the separated data and command flow that are required for initiating the DMA transmission of the received payloads to the host for \texttt{RDMA WRITEs} and \texttt{RDMA READ RESPONSEs}. Also in this stage, incoming \texttt{RDMA ACKs} are transformed into completion events that are transmitted via an additional control bus and used for flow control as described later. 

The TX part of the pipeline has two separate inputs from the host side for both data and commands. In the first step, the information received from the commands are used to potentially release the required outgoing payload data from the retransmission multiplexer (see below in \Cref{retrans_buff}). Then, the command flow is combined with the QPN-related information retrieved from the MSN table to generate the RETH and with the PSN obtained from the state table to form the BTH. 
The merging of the incoming payload from the data bus with the intermediate RETH and BTH is interleaved with these header-forming steps. 
After the formation of the core InfiniBand packet, the required UDP and IP headers are generated and pre-pended on the outgoing TX path. 
It then leads to the ICRC module for the line-rate calculation and insertion of the checksum, before the packet is forwarded to the CMAC for the generation and transmission of the Ethernet frame. 

\subsection{Retransmission Buffers and Stream Mux}
\label{retrans_buff}
The retransmission and stream mux module serves two main purposes for incoming data from the host:  
First, to avoid mutual data blocking for independent outgoing \texttt{RDMA WRITEs} and \texttt{RDMA READ RESPONSEs}, these data flows are served via separate AXI busses from the host, but need to be merged into a single stream input for the packet processing pipeline. 
Second, if retransmissions are required due to packet timeouts or PSN sequence errors, the required payload should not be requested from the host to avoid an additional PCIe transaction with the incurred latency overhead. Instead, all transmitted payloads need to be held in card memory until remote acknowledgement of reception. 

\begin{figure}[t]
    \centering
    \includegraphics[width=0.9\columnwidth]{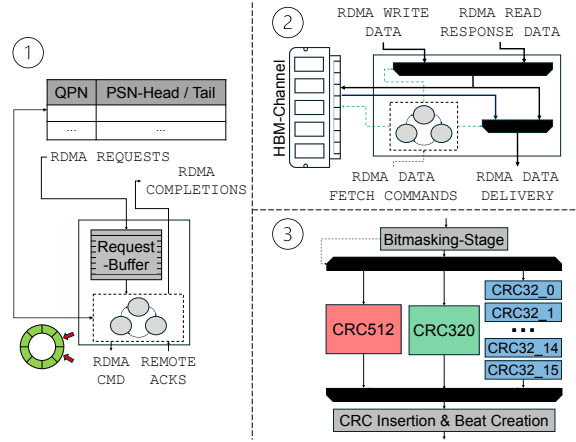}
    \caption{Building blocks of BALBOA: \ding{192} - ACK-clocked flow control, \ding{193} - Retransmission logic, \ding{194} - ICRC pipeline}
    \label{fig:Core Building Blocks}
\end{figure}

Within this module (\Cref{fig:Core Building Blocks} - \ding{193}), the two incoming busses are demultiplexed into a single stream via arbitration, while all incoming payloads are buffered in a dedicated, directly exposed HBM channel. Payloads, whether incoming from host or reloaded from HBM, are picked up and released by incoming commands from the packet processing pipeline. 

\subsection{Network Request Crediting}
Complimentary to the ability to retransmit timed-out packets, BALBOA also controls the acceptance of incoming packets based on the available processing and queueing capacity for the received payloads along the entire datapath leading to the host. This is especially important for network streams as the relatively high latency to remote nodes, compared to local streams within the network stack, leads to a larger number of outstanding requests, which in case of for example incoming \texttt{RDMA WRITEs} are also remotely issued and can therefore meet the receiving local stack in any state of available queuing capacity. 
Besides placing deep queues between the different processing cores to facilitate a large number of outstanding remote requests, BALBOA has direct access to a crediting mechanism that signals the consumption capacity of the host-facing parts of the design. In case of an insufficient credit count for the currently processed packet it is dropped to avoid stalling the entire host-facing datapath. If that happens, BALBOA relies on the retransmission capability of the remote node to receive the missing packet at a later point when the datapath has been cleared and can accept more payload.  

\subsection{Flow Control Logic}
RoCE BALBOA uses an ACK-clocked flow control that is widely portable, fits well with the timeout- and PSN-triggered retransmissions of RoCE and is compatible with the implementations in ASIC-based commodity NICs. It is extendable to support existing congestion control algorithms such as DCQCN \cite{dcqcn_congestion_control} or TIMELY \cite{timely_congestion_control}. At the same time, the utilized CMAC core on the evaluation board does support PFC and can therefore be configured to fulfill this standard as well \cite{CMAC_IP_Guide}.  
The flow control logic is implemented on the control path (\Cref{fig:Core Building Blocks}), where incoming RDMA requests can either be forwarded to the packet processing pipeline or queued to control the number of outstanding packets on the wire. The remaining budget of packets to be sent is stored in a flow control-adjacent memory per QP, decreased by passing requests and increased by incoming RDMA ACKs as indicator of completed and no longer outstanding transmissions. This design allows for easy adaptation of any required congestion control scheme, including those mentioned earlier for RDMA congestion management. 

\subsection{ICRC Pipeline}
As described in \cite{schelten_rdma_1}, the calculation of the ICRC-checksum over the completed packet is a performance-critical bottleneck and challenge for any RoCE implementation in reconfigurable hardware due to hard line rate requirements and the combinatorial complexity of the CRC32 algorithms. RoCE BALBOA reimplements the concepts presented in \cite{schelten_rdma_1}, based on the observation of common \texttt{tkeep} fill levels of AXI beats in RoCE packets, as shown in \Cref{fig:Core Building Blocks}. By providing separate, parallel pipelines for full 512bit and partial 320bit AXI beats, the partial checksums for these transmissions can be calculated in one shot with minimum latency. A third, multi-stage pipeline for chunks of 32 bits length allows to calculate the required checksum for any possible AXI beat, given the guaranteed 4 Byte alignment of any RoCE message. 

\subsection{Software Interface}
RoCE BALBOA follows a programming model very similar to traditional InfiniBand commands to ensure portability and accessibility with existing programs. If embedded in the Coyote v2 shell, the main handle for a RDMA flow through is a Coyote thread, which includes all abstractions for QP setup and management. Furthermore, it ties this RDMA-flow to a user-specified application offload residing in the NIC.
\vspace{8pt}

\begin{lstlisting}{language=C++} 
// Create a cThread as main handle
// for a single QP
coyote::cThread<std::any> 
coyote_thread(DEFAULT_VFPGA_ID, getpid(), 0);

// Obtain a RDMA buffer and exchange QPs
// with the remote node via TCP-sockets
int *mem = (int*) coyote_thread.initRDMA
(max_size, coyote::defPort, server_ip.c_str());
\end{lstlisting}

RoCE BALBOA relies on an initial exchange of QP information with the remote node via out-of-band TCP channels. As displayed above, the function to obtain a RDMA buffer includes a completely hidden abstraction for this task, so that the user directly obtains a pointer to memory that is visible to and accessible for the remote node. 
The local management of the RDMA buffer is possible via a scatter-gather element which allows to spread incoming transactions to multiple offset locations within the pre-allocated memory buffer and, respectively, read from multiple such locations for outgoing data transmissions: 
\begin{lstlisting}{language=C++}
// Obtain a scatter-gather-object for 
// buffer management and set its length
coyote::sgEntry sg;
sg.rdma = { .len = 64 };
\end{lstlisting}

Finally, also the actual network transport offloads via standard InfiniBand verbs and the polling on their completions is possible through the Coyote-thread: 

\begin{lstlisting}{language=C++}
// Invoke a RDMA-WRITE
coyote_thread.invoke
(coyote::CoyoteOper::REMOTE_RDMA_WRITE, &sg);

// Check for the number of completed 
// incoming WRITEs
coyote_thread.checkCompleted
(coyote::CoyoteOper::LOCAL_WRITE);
\end{lstlisting}

\subsection{Traffic Sniffer as a Debugging Utility}
To simplify the use of RoCE BALBOA in a production environment, we provide a fully performance-preserving traffic sniffer for bidirectional 100G network traffic on both the RX and TX path that captures all traversing packets and generates a standard PCAP file for traffic analysis and network debugging with tools such as Wireshark. Based on a header-selecting traffic filter at the CMAC level, the traffic sniffer utility can be tuned by the user to only capture packets of a certain protocol, such as RoCE v2, and to omit packet payloads to reduce the bandwidth and memory footprint caused by the instrumentation.    

\section{Compute Offloads in BALBOA}
BALBOA enables two types of compute offloads: protocol enhancements at the core of the stack to enhance network functionality, and application-specific offloads in configurable logic slots on the datapath.

\subsection{Protocol Enhancements as a Service}
The RDMA protocol enhancements are located close to the packet processing pipeline and interact closely with both data- and control flow of the network stack. In comparison to function-specific offloads, they fulfill network-specific tasks, but are still exchangeable and extensible. Function offloads as user-defined applications can utilize one or multiple of these protocol enhancements as a service of the network stack, just like the general RDMA transport mechanism. In general, two forms of protocol enhancements are possible: 

On-datapath protocol enhancements are located right before or after the packet processing pipeline and reside on the AXI streams for either network packets or payloads, while not interfering with the actual packet processing of the packet processing pipeline (\Cref{fig:wide-image}, component \ding{192}). 

Contrary to this, parallel-path protocol enhancements (\Cref{fig:wide-image}, components \ding{193}) are located next to the core packet processing pipeline, operate on a multiplexed copy of the incoming or outgoing packets and can directly interact with the packet processing logic via interfaces to the pipeline. The latency of these protocol enhancement services needs to be hidden by the partial latency of the parallel packet processing pipeline to avoid stalling the entire network stack. 

\subsubsection{AES Encryption}
We show the capabilities of on-datapath protocol enhancements along an example of AES ECB encryption.
For AES ECB, the subsequent blocks of data that are encrypted are independent, which allows for pipelining and therefore easy integration into the network datapath. 
To implement this, RoCE BALBOA deploys an open-source AES ECB IP core\footnote{https://github.com/fpgasystems/hw-acceleration-of-compression-and-crypto} (\Cref{fig:wide-image}, component \ding{192}). 
The AES keys can be exchanged as part of the out-of-band QP initiation and then stored locally at the network stack, where they can be retrieved and inserted based on the incoming RDMA commands on the control plane. 
QP state information can be retrieved easily based on the accompanying control path.
This only adds latency to the overall performance and does not compromise throughput as long as line-rate compliance of the pipeline design is guaranteed. 

\subsubsection{ML-based Deep Packet Inspection}
As an example for parallel-path protocol enhancements (\Cref{fig:wide-image}, component \ding{193}), RoCE BALBOA deploys an open source machine learning-based deep packet inspection design \cite{ml_based_dpi_for_rdma} 
To address the lack of access control in RoCE v2 through the combination of easy-to-spoof RDMA flows and direct access to application memory with OS-bypassing, this service leverages ultra low latency inference of ML on FPGAs to offload packet checking to the NIC and identify potentially malicious executables embedded in hijacked RDMA flows at line rate. 
This maintains the key performance principle of host bypassing in RDMA while marking potentially malicious packets before they even reach the host memory and can inflict damage. 
We trained the ML model on common big data payloads such as CSVs, PNGs, and TXTs versus compiled malware executables and implemented it for FPGAs with the hls4ml-compiler framework \cite{duarte_hls4ml}.
The deep packet inspection module receives every incoming payload on a datapath parallel to the packet processing pipeline. 
The end-to-end latency for inference and therefore the decision whether the received packet is acceptable or malicious occurs in 44ns per AXI beat.
This is shorter than the latency of the packet processing pipeline and therefore does not result in a performance penalty. 
Finally, the aggregated DPI decision is communicated to the packet processing pipeline and included as a flag in the resulting host-directed command. 
This allows to raise a user interrupt in software for the case of an identified malicious packet, so that the user can decide how to handle this incoming payload, i.e., by running an additional, software-based analysis tool on the data.

The tight integration of the DPI model into the core packet processing logic is a prime example of a protocol enhancement that is only possible with a fully customizable and open source design. Following the principles of this parallel-datapath service, other modules can be installed in a similar position to run for example data analytics on incoming packets and impact the further processing based on the outcome of the service function. 

\subsection{On-datapath Application-specific Offload}
Thanks to its generic AXI stream interfaces, RoCE BALBOA allows users to deploy arbitrary offloaded logic on the RX- and TX-datapath to manipulate incoming and outgoing payloads. For example, this may be used for data-preprocessing offloaded to the edge of the network. Multiple design criteria must be met when designing such offloads: 
\begin{itemize}[itemsep=2pt, parsep=0pt]
    \item For both the RX and TX direction, special consideration is required for any operation that changes the size of the data stream. Especially for data expansion operations, the network stack exerts backpressure if such an operation exceeds the available 100G bandwidth. 
    \item Similarly, all operations placed on the RX datapath for incoming packets have to operate at line rate and must not stall at any point to avoid accumulating backpressure to the network stack and thus the physical network interface as it would inevitably lead to packet drops. 
    \item Finally, the user has access to both the data and control flow for incoming packets. This allows to manipulate the local data handling, i.e., splitting up a single packet into multiple DMA transfers to different local buffers for a hardware-accelerated scatter-gather functionality. On top of that, data can also be forwarded to local NIC-attached memory for further processing or storage. 
\end{itemize}
We provide a comprehensive simulation framework that allows to test any user application against the behavior of RoCE BALBOA for an efficient development process. 

\begin{figure*}[t]
  \centering
  \begin{subcaptionbox}{Throughput evaluation for \texttt{RDMA WRITE}. \label{fig:WRITE_THR}}[0.49\linewidth]
    {\includegraphics[width=\linewidth]{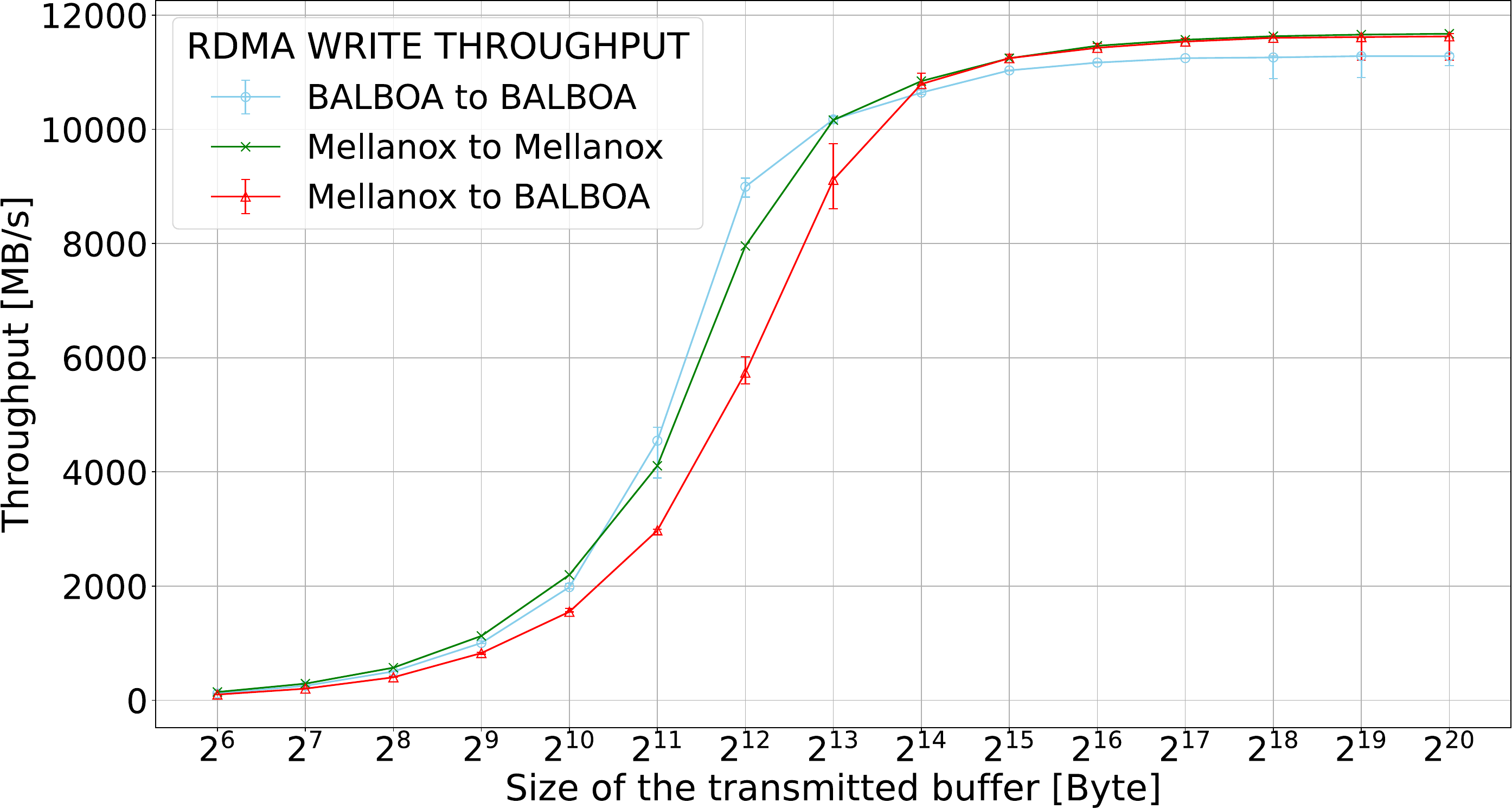}}
  \end{subcaptionbox}
  \hfill
  \begin{subcaptionbox}{Latency evaluation for \texttt{RDMA WRITE}. \label{fig:WRITE_LAT}}[0.49\linewidth]
    {\includegraphics[width=\linewidth]{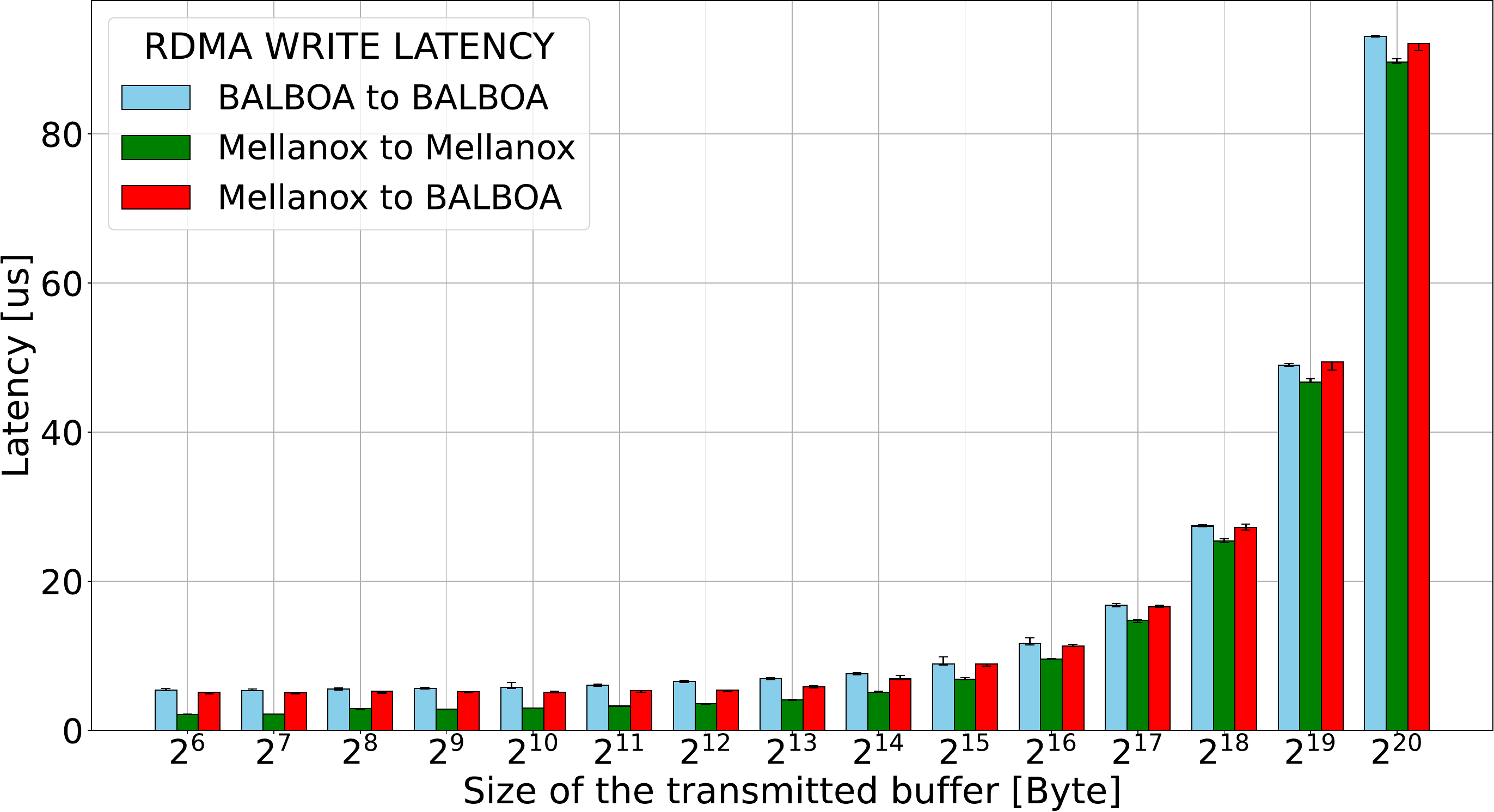}}
  \end{subcaptionbox}
  
  
  \begin{subcaptionbox}{Throughput evaluation for \texttt{RDMA READ}. \label{fig:READ_THR}}[0.49\linewidth]
    {\includegraphics[width=\linewidth]{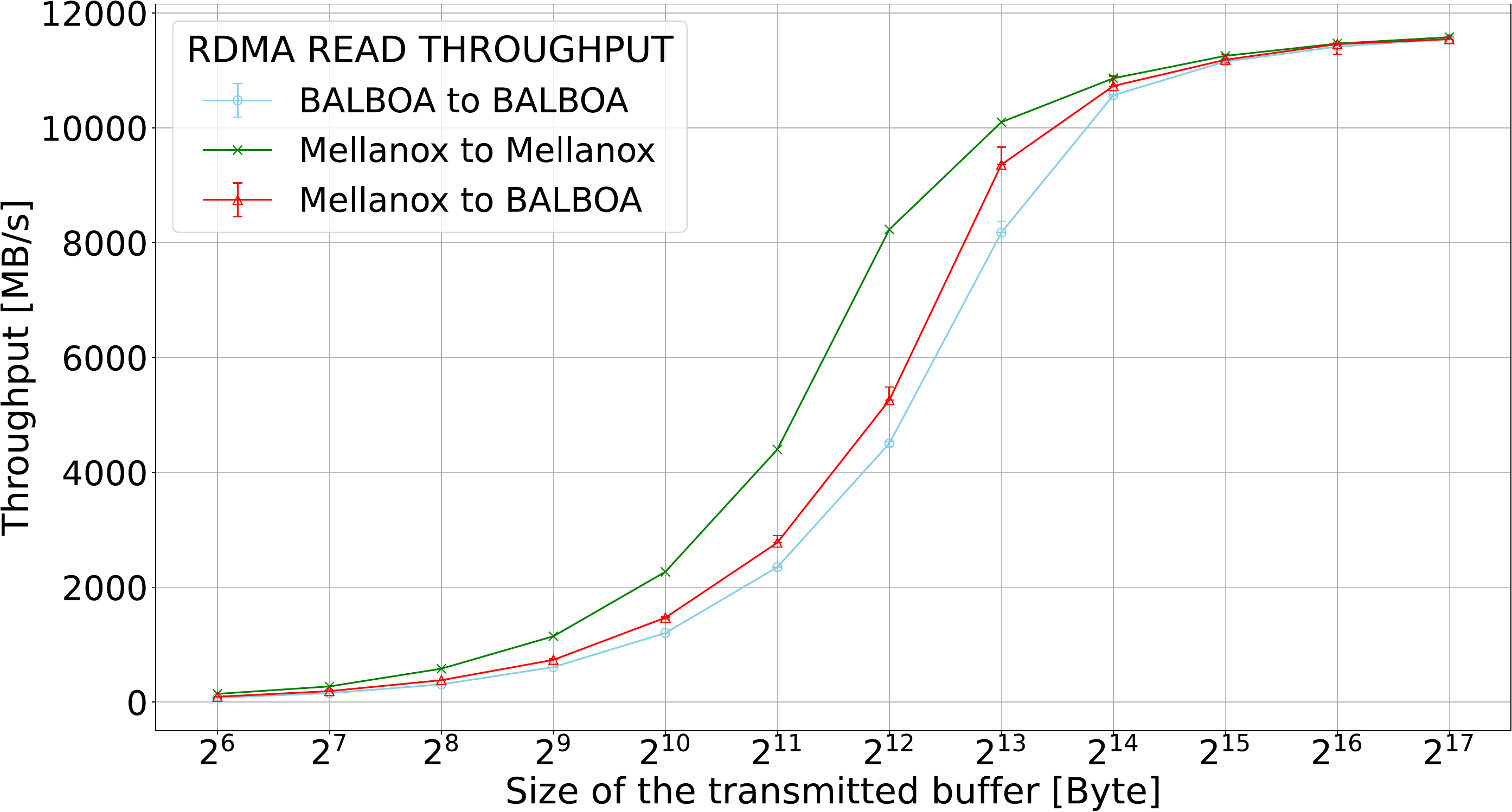}}
  \end{subcaptionbox}
  \hfill
  \begin{subcaptionbox}{Latency evaluation for \texttt{RDMA READ}. \label{fig:READ_LAT}}[0.49\linewidth]
    {\includegraphics[width=\linewidth]{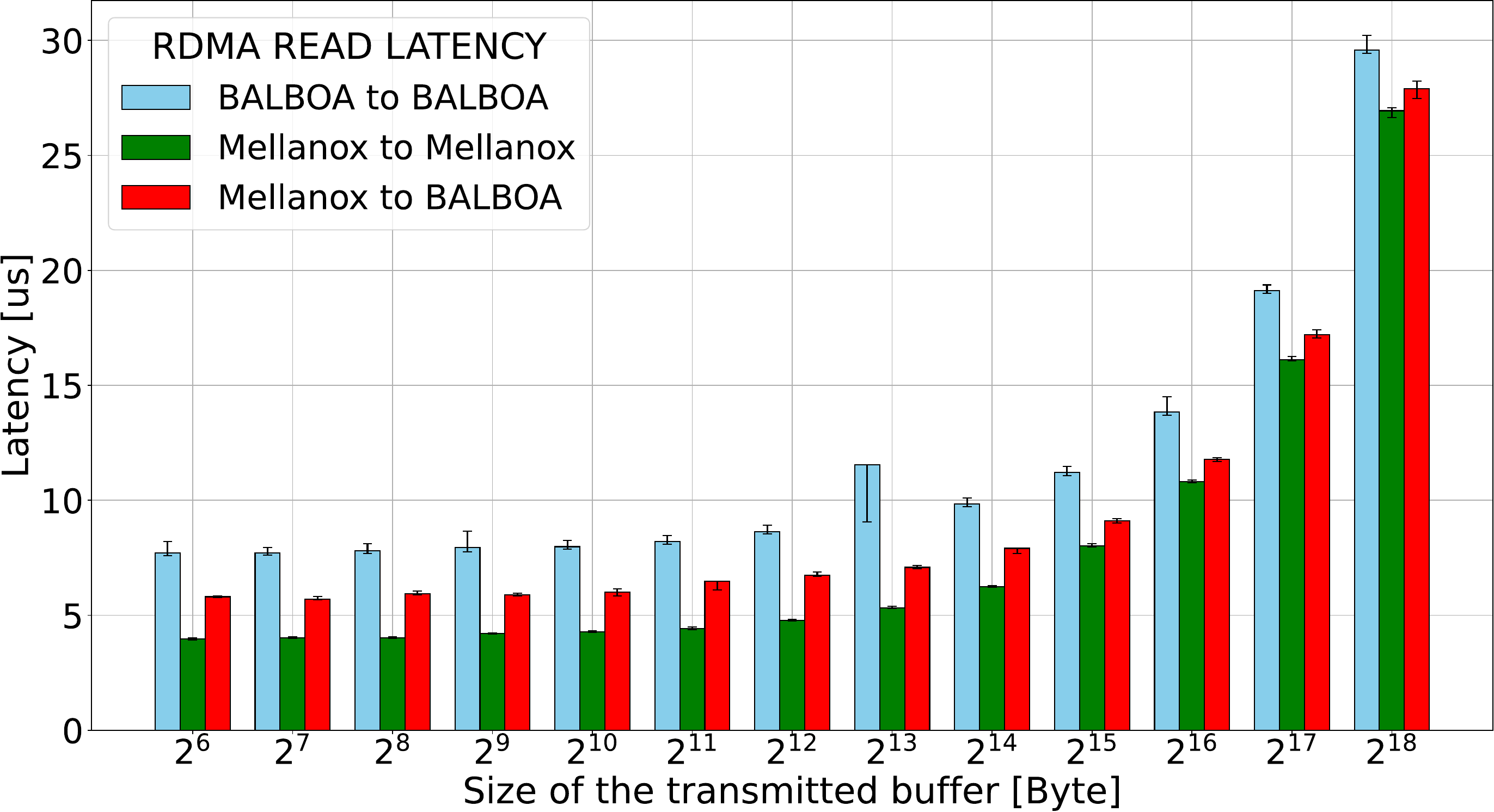}}
  \end{subcaptionbox}
  
  \caption{Performance evaluation of RoCE BALBOA in a switched 100G datacenter network. Displayed are average results over 100 repetitions as well as P5- and P95-percentiles as error bars (except for Mellanox-to-Mellanox throughput).}
  \label{fig:RDMA_microbenchmarks}
\end{figure*}

\section{Experimental Evaluation}
In the following, we evaluate the performance of RoCE BALBOA along multiple measures.
All experiments were conducted in a public research cluster
\cite{moya2023hacc} that resembles a typical datacenter network: The 100G subnet connects servers with both Mellanox ConnectX-5 commodity NICs and AMD Alveo U55C FPGA accelerators as PCIe cards over CISCO Nexus 9000 Series switches. We evaluate the RDMA performance between homogeneous and mixed combinations of RoCE BALBOA on the Alveo FPGAs and the Mellanox NICs through a single layer of switches, but verify the same functionality and throughput also in a 2-tier fat-tree network topology with leaf and spine switches, as this is more common in HPC-oriented cloud setups. For the performance optimization of the network, the maximum transmission unit (MTU) is set to 4K, and the CMAC IP utilized by RoCE BALBOA on the Alveo U55C FPGA is set up to utilize the smallest possible inter packet gap (IPG) compliant with IEEE std 802.3-2012 clause 82 \cite{ieee802_3_2012}. The deployment and complete system used for the experiments is the one shown in \Cref{fig:wide-image}.

\subsection{Basic Performance \& Cross-compatibility}
RoCE BALBOA is evaluated in the described setup for latency and throughput at various transmitted buffer sizes, both in FPGA-to-FPGA flows as well as heterogeneous FPGA-to-Mellanox QP connections (\Cref{fig:RDMA_microbenchmarks}). Additionally, Mellanox-to-Mellanox results are plotted for comparison. Following the general practice of network performance testing, latency measurements are obtained through repeated single buffer transmissions and completion-polling, while the values for throughput performance are gathered from repeated batch-transmissions of 64 buffers subsequently. 

When first comparing only the BALBOA-BALBOA performance with the Mellanox-Mellanox performance, it becomes evident that both the FPGA and the commodity NIC perform very similarly, especially for \texttt{RDMA WRITE} operations: As shown in \Cref{fig:WRITE_THR}, both designs reach the 100G saturation point for 32kB buffers. In terms of latency (\Cref{fig:WRITE_LAT}), the higher internal clock frequency of the ASIC-based NIC results in a latency advantage especially for small buffers, whereas this discrepancy becomes less impactful for larger messages. In any case, both devices have a steady latency performance with a negligible tail latency as shown by the 95 percentile lines. 
The same general observations hold truth when analyzing the \texttt{RDMA READ} behavior.
However, a more distinct performance difference for throughput performance (\Cref{fig:READ_THR}) for medium-sized buffers becomes visible. 

Heterogeneous connections between BALBOA and a Mellanox NIC exert properties of both devices as described before.
For example, the obtained latency values fall between those measured for either of the homogeneous flows. 
The measurements demonstrate that RoCE BALBOA is capable of saturating a 100G link for both \texttt{RDMA READ} and \texttt{WRITE} in connections to both another FPGA and a commodity NIC. 

\subsection{Performance Breakdown of RDMA Latency in RoCE BALBOA}
\begin{figure}[t]
    \centering
    \includegraphics[width=\columnwidth]{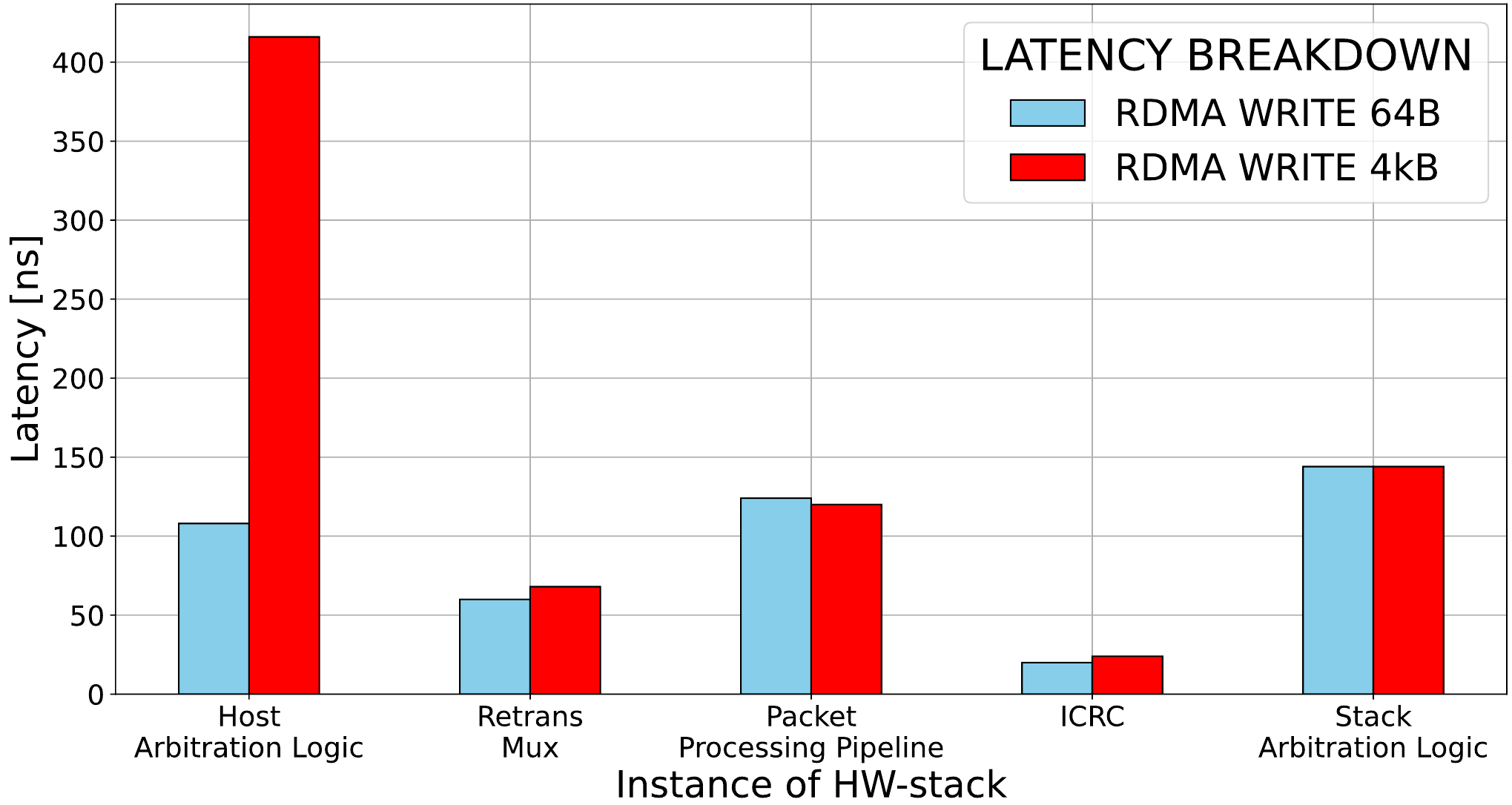}
    \caption{Breakdown of the end-to-end latency for various pipeline elements on the datapath for two sizes of outgoing \texttt{RDMA WRITE} packets.}
    \label{fig:LAT_BREAKDOWN}
\end{figure}

The performance characteristic of RoCE BALBOA can be understood by a thorough analysis of the end-to-end latency of the different building blocks of the logical design on the data and control path, measured from first-byte-in to first-byte-out. We measure for both a very small message (64 Bytes) as often used for doorbell registers and full MTU-sized messages (4~kB). 
As visible in \Cref{fig:LAT_BREAKDOWN}, the latency of the actual RDMA stack is dominated by the packet processing pipeline and not the checksum calculation, proving the efficiency of the chosen ICRC implementation as one of the key performance bottlenecks of any network stack implementation. Apart from that, the arbitration logic before and after the network stack itself has a major impact on the overall latency and shows differences depending on the transmission length.  

\subsection{Multi-QP Performance Scaling}
\begin{figure}[t]
    \centering
    \includegraphics[width=\columnwidth]{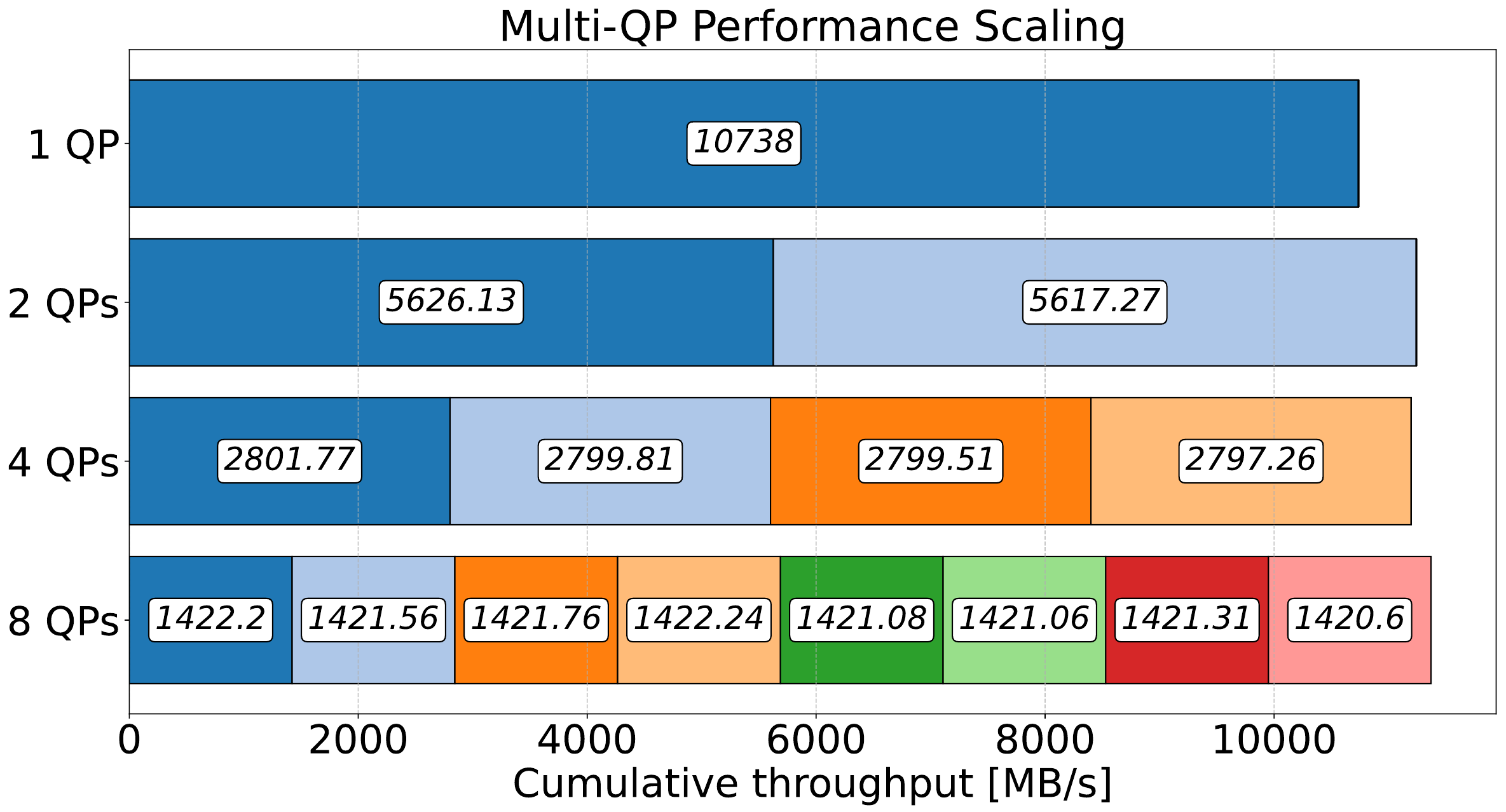}
    \caption{Bandwidth distribution across multiple QPs for 32k \texttt{RDMA READs} in batches of 64 execution over 100 repetitions.}
    \label{fig:MQP_SCALING}
\end{figure}

RoCE BALBOA allows to serve multiple QPs from independent threads, as it is an abstraction for multi-tenancy of the physical network access. By intertwining batched transmissions of multiple QPs, we are able to show that the arbiters used in RoCE BALBOA are capable of evenly distributing the available link bandwidth between the competing flows as depicted in \Cref{fig:MQP_SCALING}. It is important to note that these flows can either be routed through the same user application offload or directed to utilize different IP cores on the datapath as intended by the QP control flow. 

\subsection{Protocol Enhancements: Encryption and DPI in BALBOA and Other Platforms}
To demonstrate the benefits of NIC-offloaded service enhancements in general and specifically network stream processing for line-rate execution, we showcase the performance of the exemplarily chosen encryption and deep packet inspection and, if applicable, compare to SW-based implementations of the same functionalities on a host CPU (AMD EPYC 7302P) coupled to Mellanox ConnectX-5 NICs. While the processing steps in RoCE BALBOA reside on the datapath and are therefore executed automatically, the execution on the CPU is triggered via a remotely written doorbell register on which the receiving host polls for completion to start the processing on the main buffer. 
Additionally, we also compare with the performance obtained from a BlueField 3 DPU as the ASIC counterpart if applicable. 

\begin{figure}[t]
    \centering

    \begin{subfigure}[t]{\columnwidth}
        \centering
        \includegraphics[width=\columnwidth]{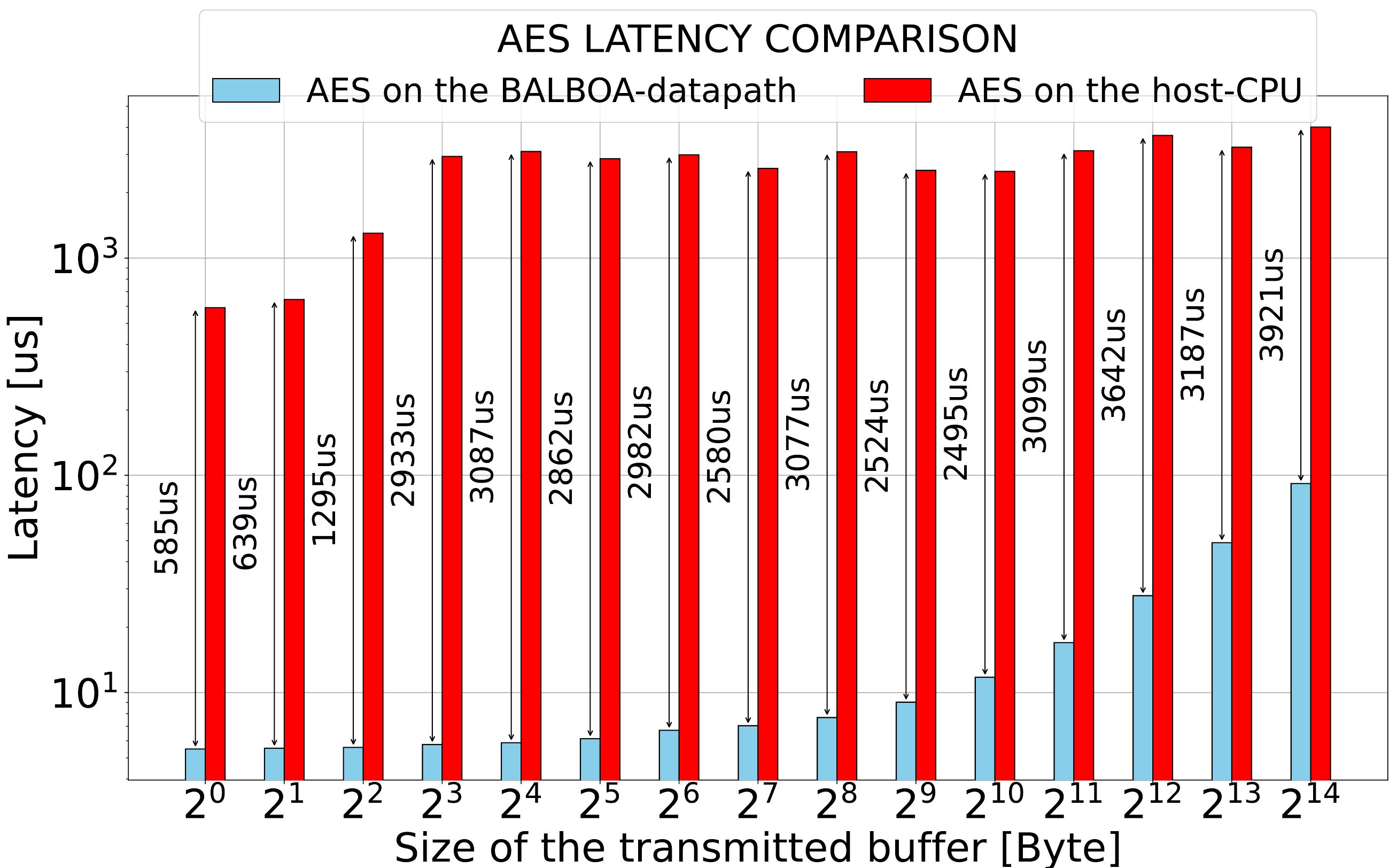}
        \caption{Latency comparison for AES on the BALBOA datapath vs. on the host CPU}
        \label{fig:aes_latency}
    \end{subfigure}

    \begin{subfigure}[t]{\columnwidth}
        \centering
        \includegraphics[width=\columnwidth]{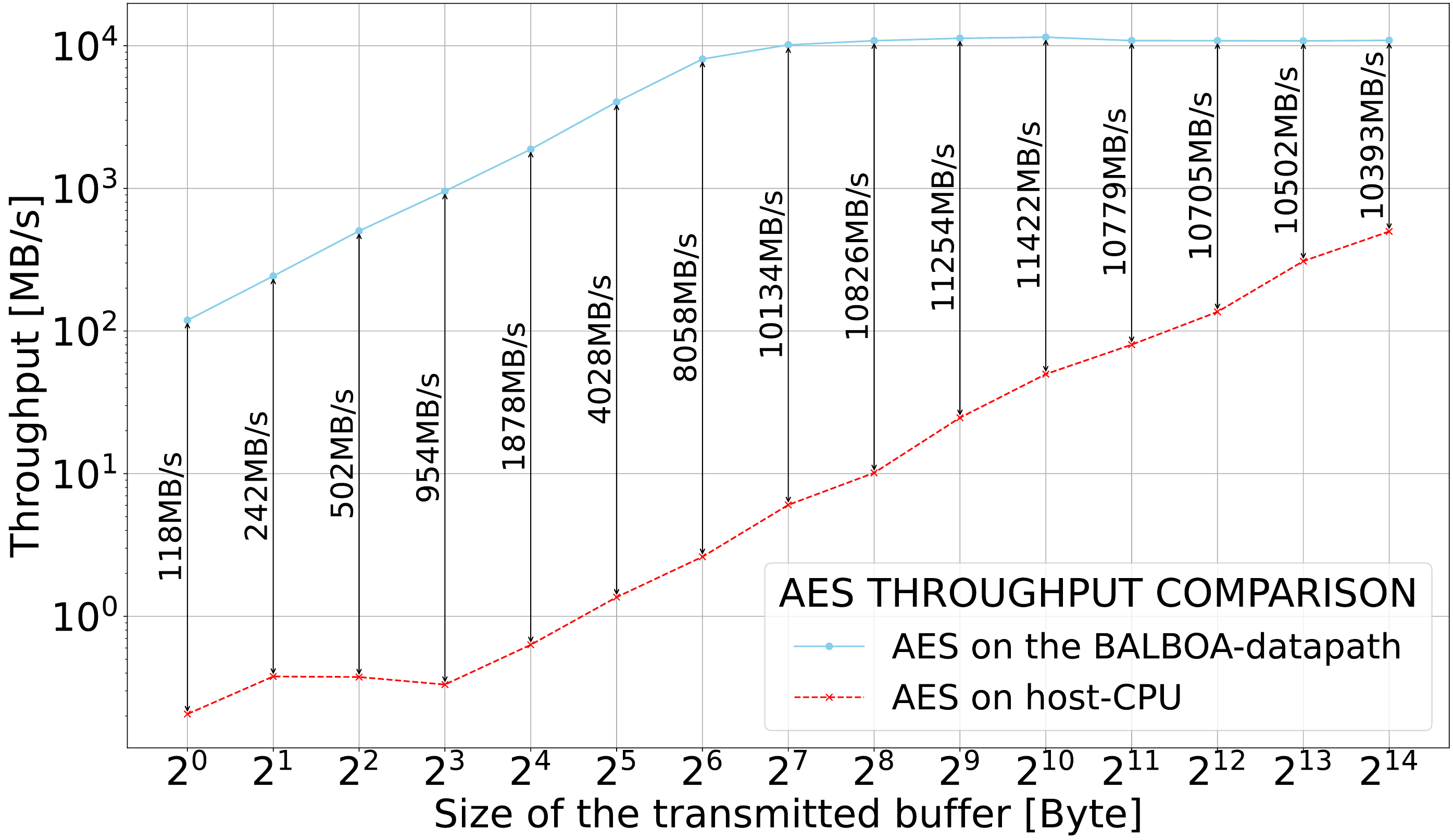}
        \caption{Throughput comparison for AES on the BALBOA datapath vs. on the host CPU}
        \label{fig:aes_throughput}
    \end{subfigure}

    \caption{Comparison of AES implementations residing on either the BALBOA-datapath or the host CPU. Y-axes in logarithmic scale for better overview and comparability.}
    \label{fig:aes_comparison}
\end{figure}

\subsubsection{AES Encryption on RDMA Traffic}
The direct comparison of latency and throughput for both the on-datapath BALBOA-implementation of AES and the realization of encryption via the host CPUs demonstrates a stark performance difference (\Cref{fig:aes_comparison}).

Even when utilizing all 16 cores of the server-grade CPU for AES encryption and decryption via the standard OpenSSL library, the software-based implementation is only able to saturate a small fraction of the 100G network bandwidth for the tested message sizes (\Cref{fig:aes_throughput}), while the hardware-based implementation in BALBOA achieves the expected throughput saturation and only adds negligible additional latency: the AES pipeline has an end-to-end processing time of 10 clock cycles or 25 ns respectively (\Cref{fig:aes_latency}). 

Omitted in the result plots, we do see that the CPU-based implementation reaches a throughput of around 3600 MB/s for gigabyte-sized RDMA buffers. While these numbers still not match the performance obtainable with the NIC offload, it becomes evident that a part of the performance gap is due to the triggering mechanism based on doorbell polling. Recent work on specialized RDMA-protocols for security and privacy~\cite{sRDMA_taranov} solves this problem partially, but also only achieves a line rate performance of 25 Gbps. This however highlights yet another performance advantage enabled by on-datapath logic, where scheduling of execution is a non-existing problem. 
The general performance difference is backed up by existing research on accelerator-based encryption, where accelerator solutions consistently outperform CPU-based solutions \cite{sap_encryption_compression}, while also not adding to the "datacenter tax" of continuously thrashing CPU-cores for repetitive tasks. 

These obvious advantages are the reason why cryptographic acceleration blocks are becoming a commodity in ASIC-based networking platforms. However, our own initial research and existing literature indicates that even those hardened encryption and decryption blocks on BlueField-cards is not able to saturate the full network bandwidth achievable with these NICs \cite{Bluefield_AES_performance}, stressing again the performance advantages possible with deeply pipelined customized solutions in reconfigurable fabric, coupled to BALBOA. 

\subsubsection{Deep Packet Inspection}
When analyzing the efficiency and performance of the DPI-service combined with BALBOA, two properties are of relevance. To begin with, the deployed fully-connected ternary neural network provides excellent performance in detecting potentially malicious executables in RDMA-packets: If whole packets contain only such unwanted payloads, the detection rate is at 97.83\%, while partially embedded executables in larger messages are still recognized with 89.35\% probability. Compared to the native false positive rate of flagging acceptable payloads of the model, a highly effective and fine-grained differentiation policy is possible based on the ML-decisions. 

\begin{figure}[t]
    \centering
    \includegraphics[width=\columnwidth]{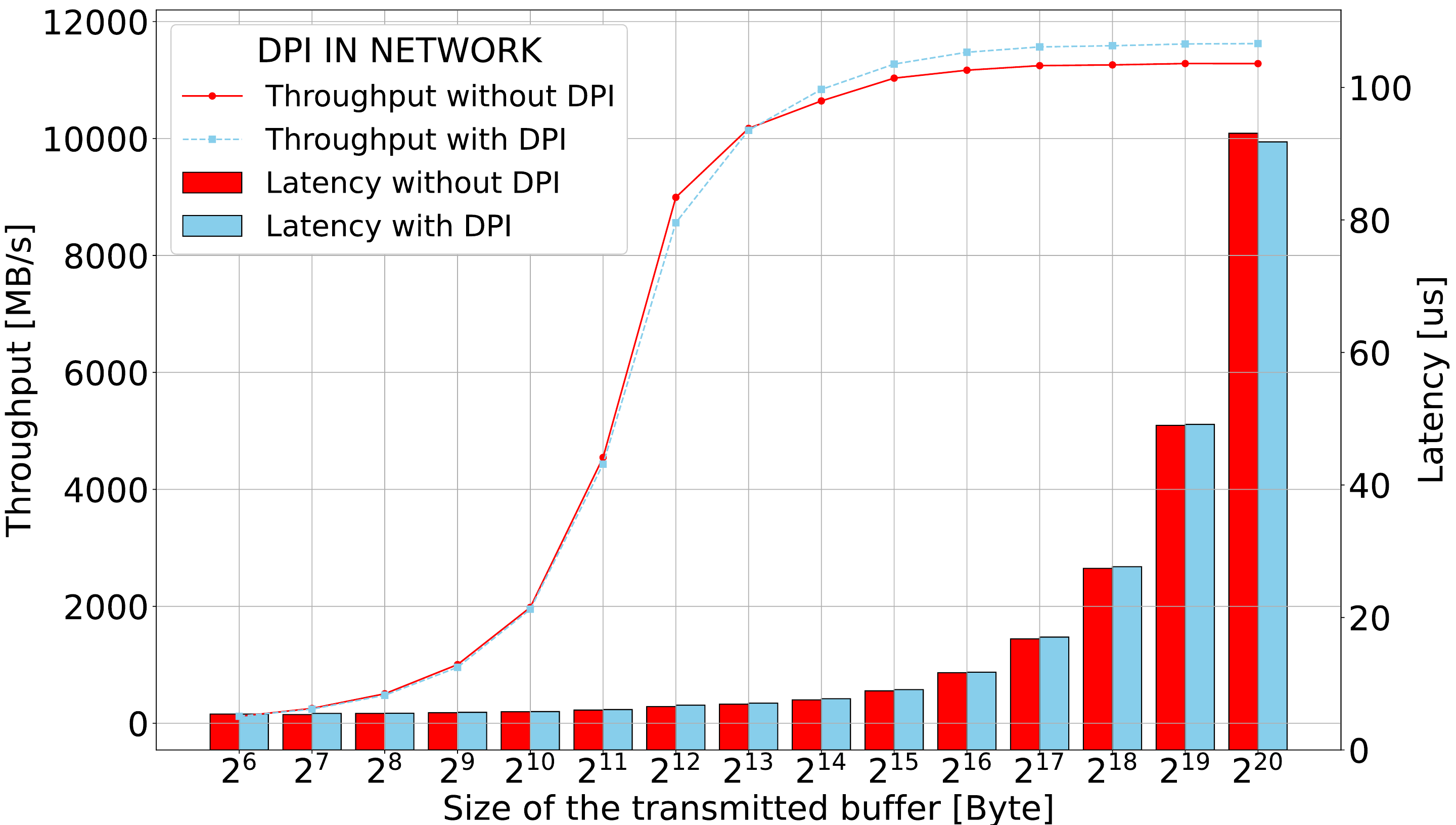}
    \caption{Throughput and latency 
    with and without deployed ML model for deep packet inspection.}
    \label{fig:DPI_PERF}
\end{figure}

At the same time, \Cref{fig:DPI_PERF} clearly indicates that the deployment of the ML model alongside the BALBOA stack does not have any negative impact on either end-to-end latency or throughput of the packet processing, as the inference latency of the service is completely hidden by the parallel packet processing pipeline. Smaller deviations of the performance curves are caused by the naturally changing conditions in the public network used for the experiments.

\section{Hardware Resource Evaluation}
Resource utilization is evaluated with post-routing results for an AMD Alveo U55C accelerator platform (\Cref{table:resource_overview}). The hardware footprint of the entire BALBOA-stack is very small compared to the available resources on a modern datacenter chip, stressing its efficiency and portability for all kind of existing NIC and accelerator designs. This rather low resource consumption is also one of the main reasons why service enhancements such as the demonstrated logic blocks for AES encrypion or ML DPI can easily be added to the network design without compromising on the available design space for application-specific function offloads.

\begin{table}[t]
\caption{Resource breakdown for RoCE BALBOA and its components, and for the AES and DPI service offloads.}
\footnotesize
\begin{tabular}{llll}
\hline
                                    & LUTs    & BRAM       & FFs           \\ \hline
RoCE BALBOA                         & 43732 [3.4\%]  & 101 [5.1\%] & 102988 [4\%]   \\
\textit{ICRC}                       & 12462 [0.96\%] & 0 [0\%]     & 23942 [0.9\%]  \\
\textit{Retrans-Logic}       & 587 [0.045\%]  & 34 [1.7\%]  & 1157 [0.04\%]  \\
\textit{Flow Control}         & 787 [0.06\%]   & 1 [0.05\%]  & 2836 [0.1\%]   \\
\textit{Packet Processing} & 29897 [2.3\%]  & 66 [3.4\%]  & 75053 [2.9\%]  \\
\textit{- Connection Table}         & 64 [0.005\%]   & 4 [0.2\%]   & 230 [0.009\%]  \\
\textit{- State Table}              & 202 [0.02\%]   & 6 [0.3\%]   & 265 [0.01\%]   \\
\textit{- Transport Timer}          & 209 [0.02\%]   & 2 [0.1\%]   & 248 [0.01\%]   \\
\textit{- MSN Table}                & 92 [0.01\%]    & 5 [0.25\%]  & 373 [0.01\%]   \\ \hline
AES Cryptography         & 65662 [5\%]    & 0 [0\%]     & 11669 [0.45\%] \\ \hline
ML-based DPI     & 54404 [4.2\%]  & 0 [0\%]     & 40508 [1.6\%]  \\
\textit{Payload Extractor}          & 6252 [0.5\%]   & 0 [0\%]     & 1135 [0.04\%]  \\
\textit{ML-DPI Decider}             & 48152 [3.7\%]  & 0 [0\%]     & 39373 [1.5\%] 
\end{tabular}
\label{table:resource_overview}
\end{table}

\section{Use Case Example: ML Preprocessing Pipelines for RDMA-to-GPU}
Deep learning recommendation models (DLRM) comprise a large fraction of all ML workloads in modern data centers and cloud computing \cite{preproc_impact}. To maintain accuracy of the recommendations while facing data drift and evolution, frequent online re-trainings are necessary \cite{bother2025modyn, grubhub_retraining}. While the input data is multi-modal and consist of images, audio or text, the training process itself operates on vector embeddings \cite{embeddings_for_recommender_models}, so a preprocessing step for conversion is required. Traditionally, such training systems utilize a hybrid hardware configuration with CPUs and GPUs \cite{hybrid_training_systems_recommender_models}. When transferring data from disaggregated storage to the GPUs for online-training of DLRMs, data preprocessing becomes a critical bottleneck, as more and more CPU-cores are required to saturate the increasing bandwidth of rapidly evolving GPUs, leading to more than 60\% share of total power consumption in DLRMs for just this initial data preparation step \cite{DLRM_preprocessing_study, Meta_Engineering_Report}. This leaves potential for an accelerated approach which combines network and compute and utilizes the capabilities of BALBOA: We demonstrate how to drastically increase efficiency and processing speed of DLRM data preprocessing by offloading it onto the BALBOA datapath in deep pipelines operating at line rate and combining it with the direct-to-GPU capability for direct forwarding to the accelerator-memory. By choosing this approach, the CPU as a potential bottleneck in the system is completely bypassed.  

\subsection{Functional and Design Overview}
For a realistic use-case scenario, we deploy operators from Meta's training toolkit \cite{Meta_Preprocessing_Pipelines, Meta_DNN_Recommendations} implemented in deep pipelines for non-stalling line-rate execution in hardware \cite{yu_preprocessing_pipelines}. The purpose of all DLRM operators is to enhance the quality of received data and extract unique features, depending on the sparsity of the input. Dense features consist of well-defined datapoints - such as prices or rating scores - and require normalization for faster model convergence, while sparse features, for example verbal classifications, need to be destilled into a more dense, binary representation. We select the following three stateless operators and concatenate them to a comprehensive preprocessing pipeline, as they have recently gained attention in research for optimization in hardware \cite{presto_preprocessing} and software \cite{goldminer_preprocessing}: 
\begin{itemize}[itemsep=2pt, parsep=0pt]
    \item \texttt{Neg2Zero}: Negative dense input features are clipped to zero. 
    \item \texttt{Logarithm}: Applies logarithm on dense input features to handle very large values efficiently in following processing steps. 
    \item \texttt{Modulus}: Restricts the range of sparse feature values. 
\end{itemize}
Since all operators have an initiation interval (II) of one clock cycle and operate on streaming busses of 64 Bytes width, pipelines composed of these operators are able to maintain the required throughput for network speed and can be placed directly on the datapath as facilitated by the BALBOA design without leading to a bottleneck for the end-to-end design. 

\begin{figure}[t]
    \centering
    \includegraphics[width=\columnwidth]{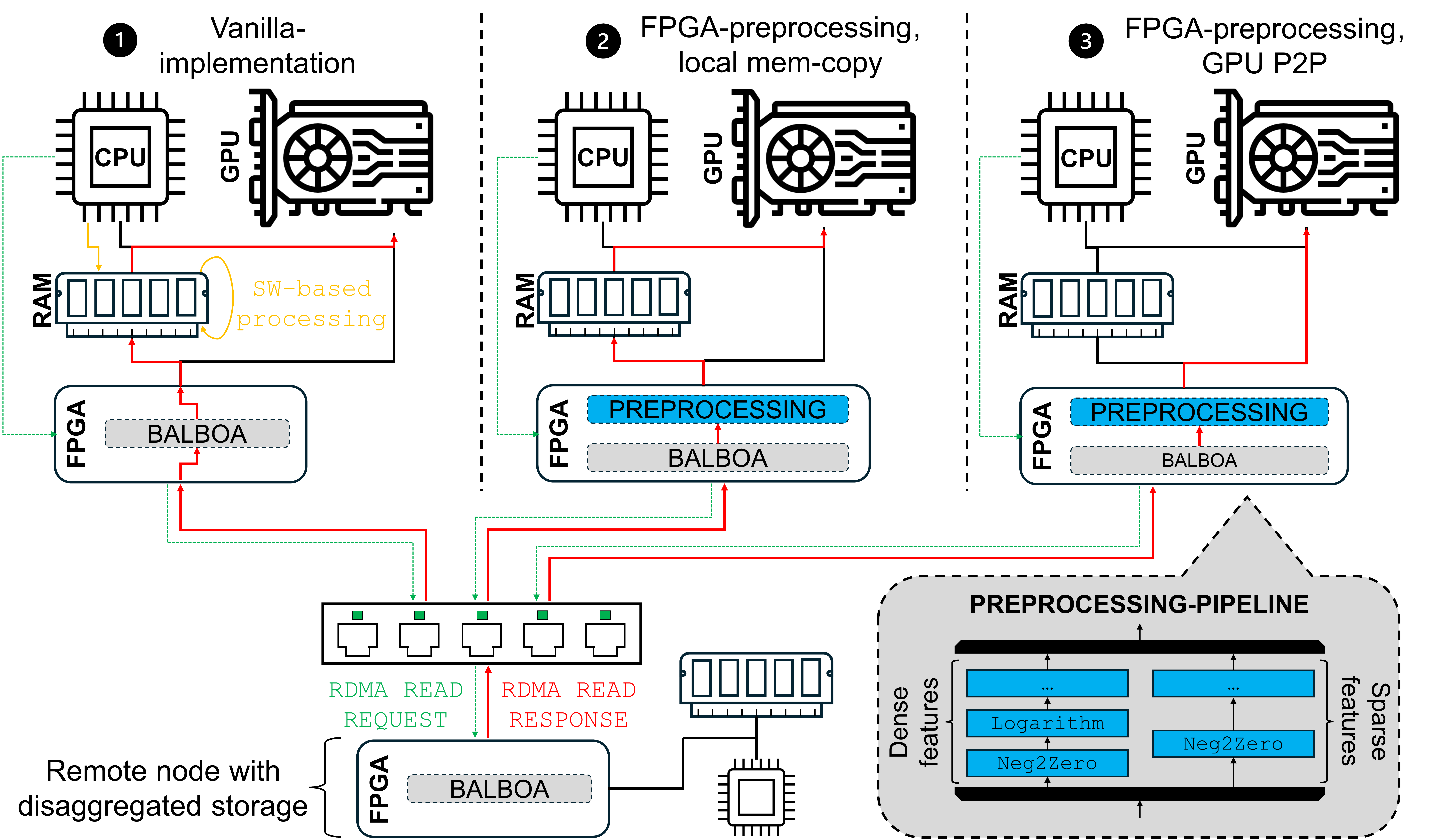}
    \caption{Visualization of the three described testing setups (\ding{192} - implementation with CPU-based preprocessing, \ding{193} - preprocessing on the FPGA, memory transport via a local buffer copy, \ding{194} - full BALBOA implementation with preprocessing on the FPGA and direct data transport to the GPU) and schematic of the implemented preprocessing pipeline.}
    \label{fig:ML_PREPROCESSING_EXPERIMENTS}
\end{figure}

\subsection{Performance Benchmarking vs. CPU Implementation}
To demonstrate both the performance advantage enabled by the on-datapath offload of the preprocessing pipeline and the direct memory access to the PCIe-attached GPU, we conduct three experiments (\Cref{fig:ML_PREPROCESSING_EXPERIMENTS}): The local node reads the input data from remote disaggregated storage via a \texttt{RDMA READ} operation. In the vanilla NIC-setup without processing capabilities or direct-to-GPU feature, the received data is written to a local buffer in host memory, preprocessed by a CPU-implementation of the logic running on a dedicated core and then locally copied to GPU-memory. In a more optimized setup, the preprocessing is allocated on the datapath of the SmartNIC while still relying on local buffer-copies from host to GPU-memory. The final setup utilizes all BALBOA-features for on-datapath preprocessing and GPU-communication. Again, all experiments are conducted in the same public research cluster with Alveo U55C accelerator cards hosting BALBOA, AMD MI210 GPU-accelerators and AMD EPYC 7V13 CPUs as host nodes. It is important to note that the FPGA and GPU are located at different NUMA-nodes in the test servers, so that the maximum possible throughput between these devices within one node is limited to approximately 70 Gbps (8500 MB/s) due to the PCIe-switch on the local datapath. 

\begin{figure}[t]
    \centering
    \includegraphics[width=\columnwidth]{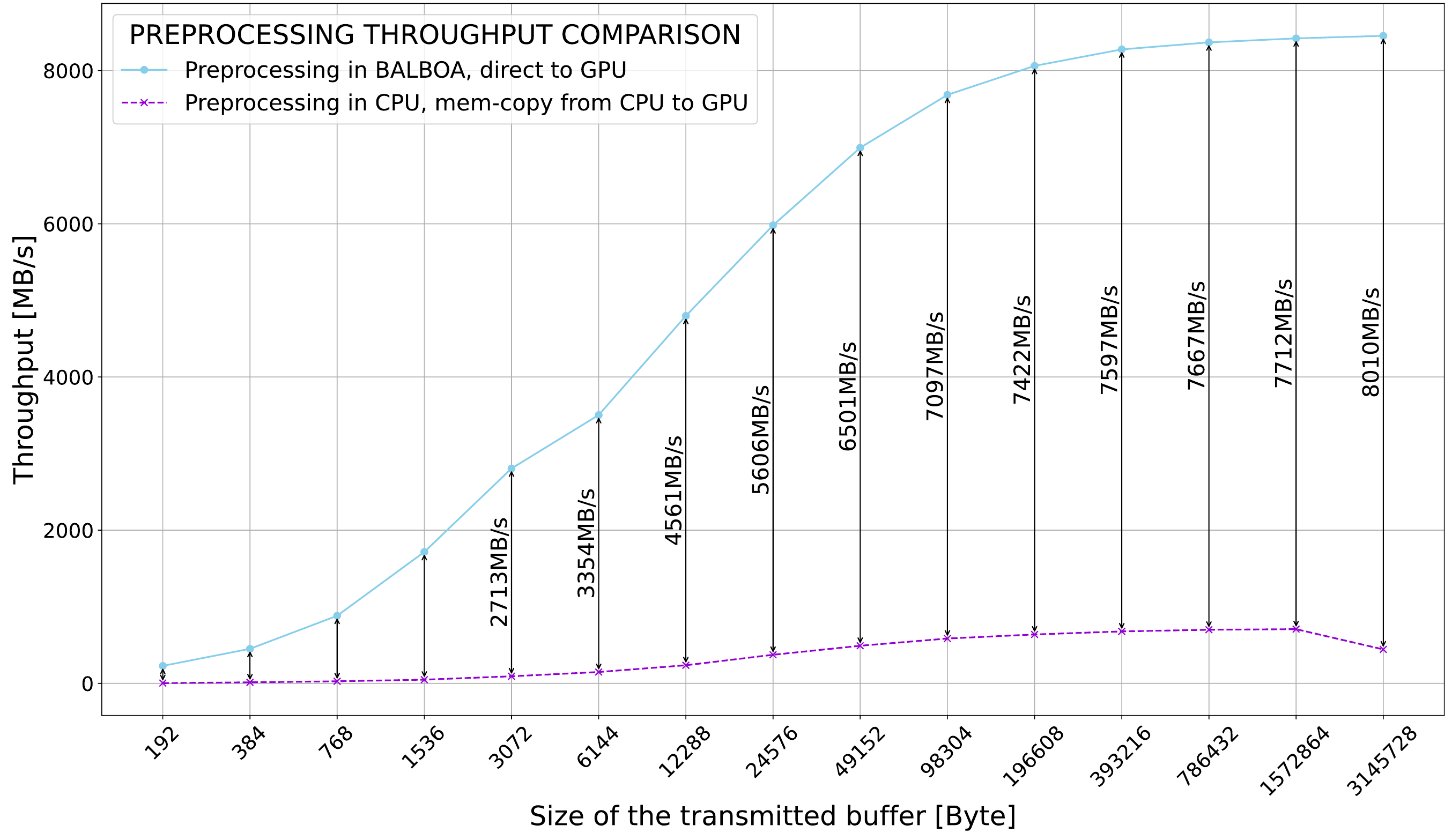}
    \caption{Throughput of ML preprocessing in BALBOA with direct-to-GPU vs. preprocessing in CPU and a local mem-copy from CPU to GPU memory.}
    \label{fig:PREPROC_THR}
\end{figure}

Comparing the vanilla implementation to the BALBOA-optimized design clearly reveals the throughput benefit unleashed by specialized, deeply pipelined on-datapath accelerator logic: Not only limits the SW-implementation of the preprocessing logic the achievable throughput to around 700 MB/s compared to the FPGA line rate speed of 8500 MB/s (\Cref{fig:PREPROC_THR}), it also effectively occupies valuable CPU cycles and adds to the data center tax. 

\begin{figure}[t]
    \centering
    \includegraphics[width=\columnwidth]{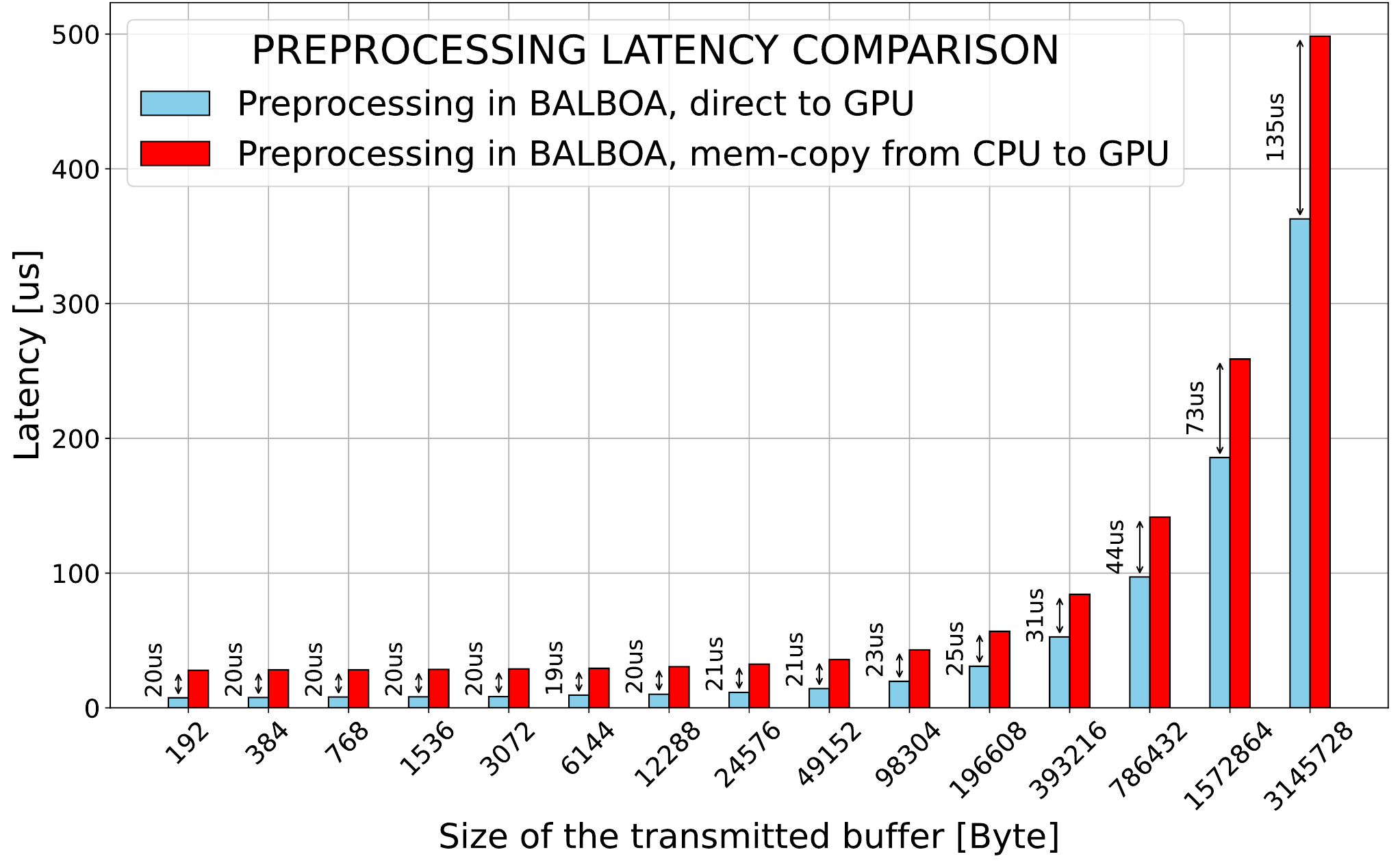}
    \caption{Latency of ML preprocessing in BALBOA with direct-to-GPU vs. an additional step through CPU memory.}
    \label{fig:PREPROC_LAT}
\end{figure}

At the same time, the latency comparison between the GPU-direct utilization of BALBOA vs. the additional step through the host memory highlights the achievable time savings on the direct path at 20-135 us (\Cref{fig:PREPROC_LAT}). Additionally, the direct data movement to GPU does not require any CPU-involvement and is therefore following the general idea of the RDMA-protocol that made it relevant for cloud computing in the first place. 

Summarizing these findings, it becomes clear that the combination of DMA-to-GPU mechanisms as often used in modern DPU-GPU setups with capable on-datapath computation in the NIC holds great promises for data-intensive preprocessing offloads to saturate the ever-increasing bandwidth of modern GPU accelerators. This helps to avoid running into CPU bottlenecks and using valuable core capacity. The use case showcases the advantages of an open RDMA stack like BALBOA and the possibilities it offers for more efficient system designs.

\section{Conclusion and Future Work}
In this paper, we present RoCE BALBOA, an open-sourced, RoCE-v2 capable and datacenter network-ready RDMA-stack for exploring and developing in-network accelerators and SmartNICs that allows for far-reaching customization with protocol-enhancing services and on-datapath acceleration of function offloads. We demonstrate the protocol enhancements with two examples: AES encryption and ML-based deep packet inspection. We also showcase a realistic use-case scenario of on-datapath function offloads for real-world ML-preprocessing pipelines. 
In future work, we will extend RoCE BALBOA as a platform for network research by implementing, e.g., congestion control and load balancing algorithms, as well as additional use-case scenarios with scale-out, network-dependent applications. For instance, we are currently working on line-rate compression and decompression logic cores to reduce the bandwidth consumption of RDMA flows. We are also combining encryption, compression, and data parsing on the datapath to increase the efficiency of data analytics. 

\section{Acknowledgements}
We would like to thank AMD for the donation of the Heterogeneous Accelerated Compute Cluster (HACC) which was  used for the development and testing of this project. 



{\footnotesize \bibliographystyle{acm}
\bibliography{sample}}

\end{document}